\documentclass[letterpaper]{article} 

\usepackage{aaai20}  
\usepackage{times}  
\usepackage{helvet} 
\usepackage{courier}  
\usepackage[hyphens]{url}  
\usepackage{graphicx} 
\usepackage{graphicx}  
\title{Bio-Inspired Adversarial Attack Against Deep Neural Networks}
\author{Bowei Xi \\
Department of Statistics \\ Purdue University \\ xbw@purdue.edu \\
\And Yujie Chen \\ Department of Statistics\\ University of
Chicago\\ chenyujie@galton.uchicago.edu \\
\And Fan Fei, Zhan Tu, Xinyan Deng \\
School of Mechanical Engineering\\ Purdue University \\ 
feif,tu17,xdeng@purdue.edu}

\begin{document}
\maketitle

\begin{abstract}
The paper develops a new adversarial attack against deep neural
networks (DNN), based on applying bio-inspired design to moving 
physical objects. To the best of our knowledge, this is the first
work to introduce physical attacks with a moving object.
Instead of following the dominating attack strategy
in the existing literature, i.e., to introduce minor perturbations to a
digital input or a stationary physical object, 
we show two new successful attack
strategies in this paper. We show by superimposing
several patterns onto one physical object, a DNN becomes confused and
picks one of the patterns to assign a class label. Our experiment
with three flapping wing robots demonstrates the
possibility of developing an adversarial camouflage to cause a targeted
mistake by DNN. We also show certain motion can reduce the
dependency among consecutive frames in a video and make an
object detector ``blind'', i.e., not able to detect an object
exists in the video. Hence in a successful physical attack against DNN,
targeted motion against the system should also be considered.    
\end{abstract}

\section{Introduction}
\label{sec:intro}

Current generation of
artificial intelligence (AI) has been very successful at complex tasks, such
as image classification, object recognition, or playing the board
game Go. Unfortunately such forward thinking AI is not secured
against potential digital and physical attacks. This paper
aims to develop bio-inspired adversarial attack using moving
physical objects, which has not been considered so far.  

Existing digital and physical attacks focus on adding minor
perturbations to the clean samples/objects to fool DNN. 
The digital attack algorithms mostly focus on generating digital adversarial
perturbations by solving an optimization problem or using a generative
model against one specific DNN,
e.g., \cite{nips2017competition}. Digital attacks are further
divided into whitebox attacks, graybox attacks, and blackbox
attacks, based on adversary's knowledge about the target DNN. 
For example, Fast Gradient Sign Method
(FGSM) \cite{advl-train-2015}, a one-step whitebox
attack, used the sign of the gradient of the cost
function  to generate adversarial perturbations. 
Carlini and Wagner (C\&W) attack \cite{attack-carliniL2-2017} solved a box constrained
optimization problem to generate adversarial perturbations. Many
digital attacks have been developed during the past few years. Recent survey
papers,
e.g., \cite{survey-attack-defense-DNN-2019,advl-10year-2018},
provided a list of the digital attacks. 

Compared with digital attack algorithms, there are fewer
work on designing physical objects to break DNN. 
\cite{attack-face-2016,attack-BIM-physical-2017,stop-sign-2018,3d-turtle-2018} 
are four examples. \cite{attack-face-2016} designed eyeglass frames to allow a
person to avoid face detection. \cite{attack-BIM-physical-2017}
printed out the digital adversarial images targeted on Inception
v3 classifier, and took a photo of the
printouts using a cellphone. They showed the cropped photo images
were also misclassified by Inception v3 classifier. 
\cite{stop-sign-2018} showed that 
placing a few stickers on a stop sign can cause it to be
misclassified as other traffic signs. \cite{3d-turtle-2018} created a
3D printed turtle that is misclassified as a rifle, by adding a
few color stripes on the shell. These work
also took the optimization approach with one specific DNN in
the optimization set-up, similar to the digital attack algorithms.  
Meanwhile they designed stationary physical objects to fool a target DNN.  

These attacks demonstrate a certain type of inherent vulnerability
in DNN. It is important to understand the vulnerabilities in DNN
in order to robustify DNN. To the best of
our knowledge, this paper is the first work to design
adversarial attack using moving physical objects to fool DNNs. 
Through our bio-inspired adversarial
attack, we show there exist more 
vulnerabilities in DNN, which suggests robustifying DNN is a more
difficult task than currently believed.   

\section{Bio-Inspired Adversarial Attack}
\label{sec:bio}

\begin{figure}[t]
\centering{
\includegraphics[width=1.1in]{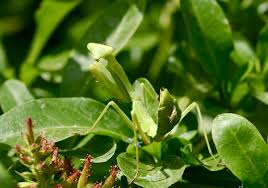} 
\hspace{0.1in}
\includegraphics[width=1.64in]{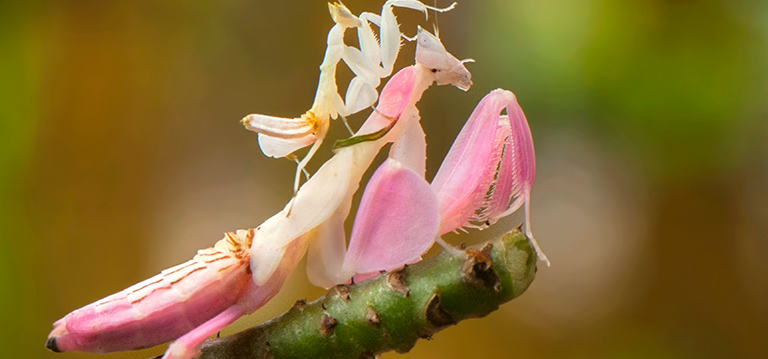}
}
\caption{Praying Mantises as Leaf (left) and Flower (right)} 
\label{mantis}
\end{figure}

Biological intelligence, not limited to human cognitive reasoning,
covers a broad range of mechanisms to make living organisms
hidden from predators and prey, and adapt to changing
environments. Figure~\ref{mantis} shows hidden praying mantises
on plants. Praying mantis stay hidden due to their camouflage
coloration. When they move, they can still remain hidden because
of the way they move. Praying mantis
coloration changes with 
the surroundings: those from dry areas are brown, whereas those
from wet areas are green. They remain motionless for a long time
until a prey gets close. When
they do move, they move with a rocking motion mimicking a swaying
plant in the wind. Learning from biological
intelligence \cite{bioAI}, we demonstrate in this paper there are more powerful
attacks against DNNs using camouflage, not limited to only adding minor
perturbations on 
digital inputs or physical objects. Hence the current generation of AI needs to
significantly improve its capabilities to face different types of
attacks. Next we show our bio-inspired adversarial
attack against DNN.  

\noindent
\paragraph{Attack by Superposition of Multiple Patterns}

\begin{figure}[h!]
\centering{
\includegraphics[width=1.2in]{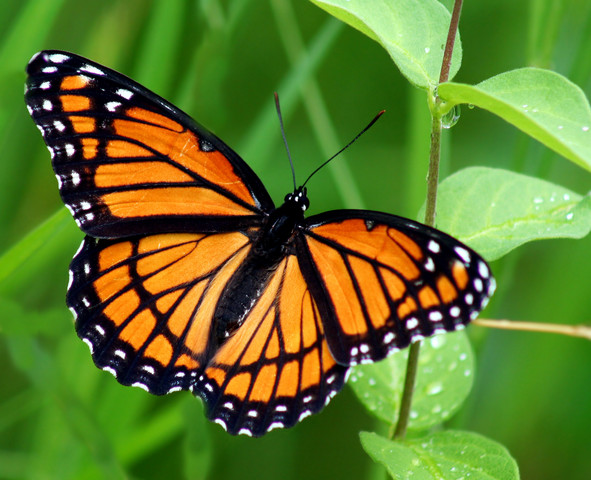} 
\hspace{0.1in}
\includegraphics[width=1.75in]{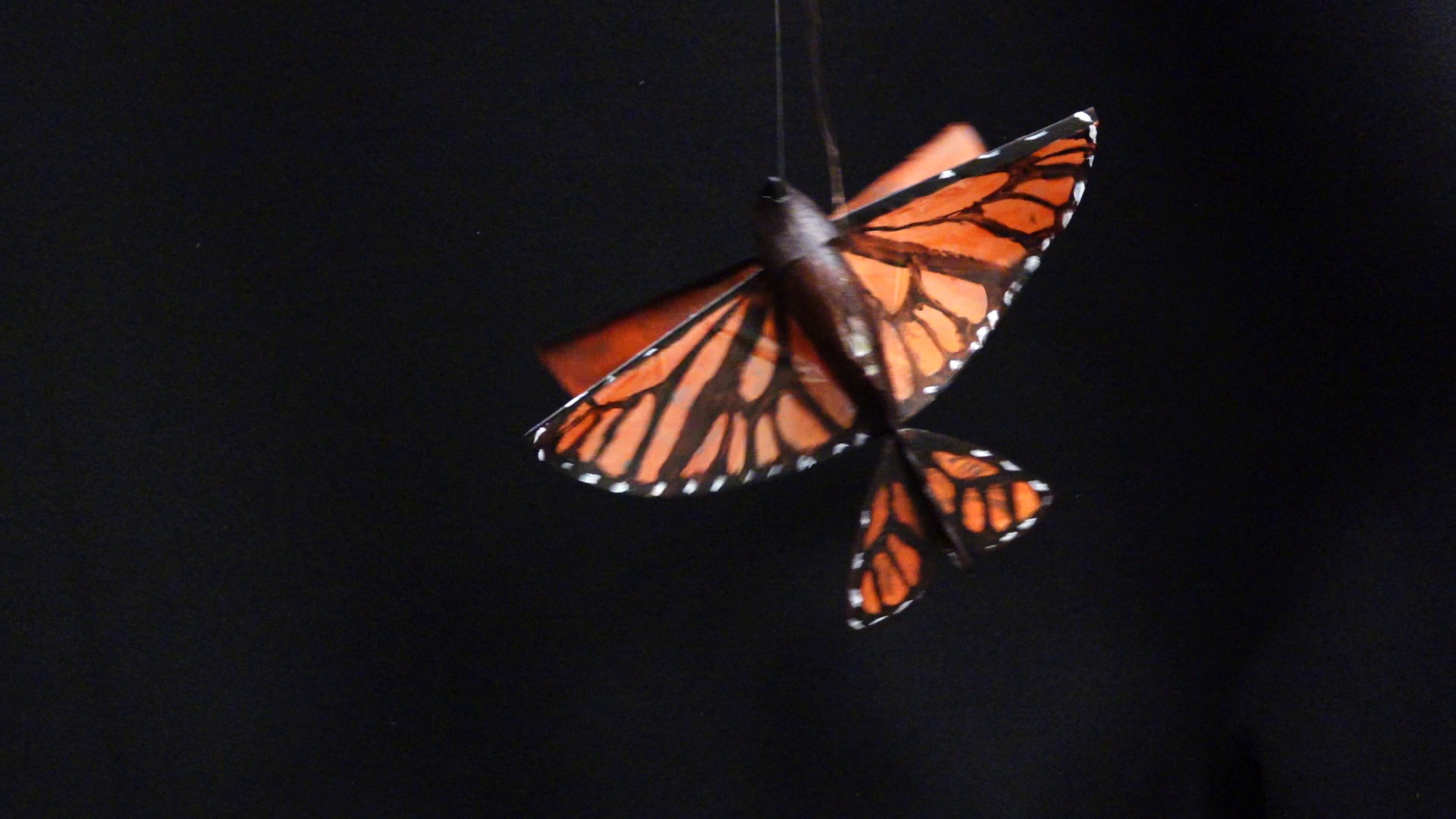}
}
\caption{Real Butterfly (left) and Robot (right)} 
\label{super-orange}
\end{figure}
\begin{figure}[h!]
\centering{
\includegraphics[width=0.84in]{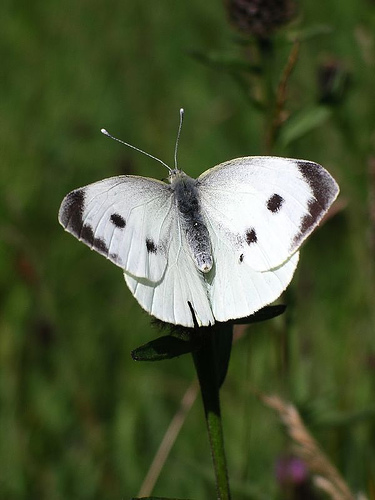} 
\hspace{0.1in}
\includegraphics[width=1.98in]{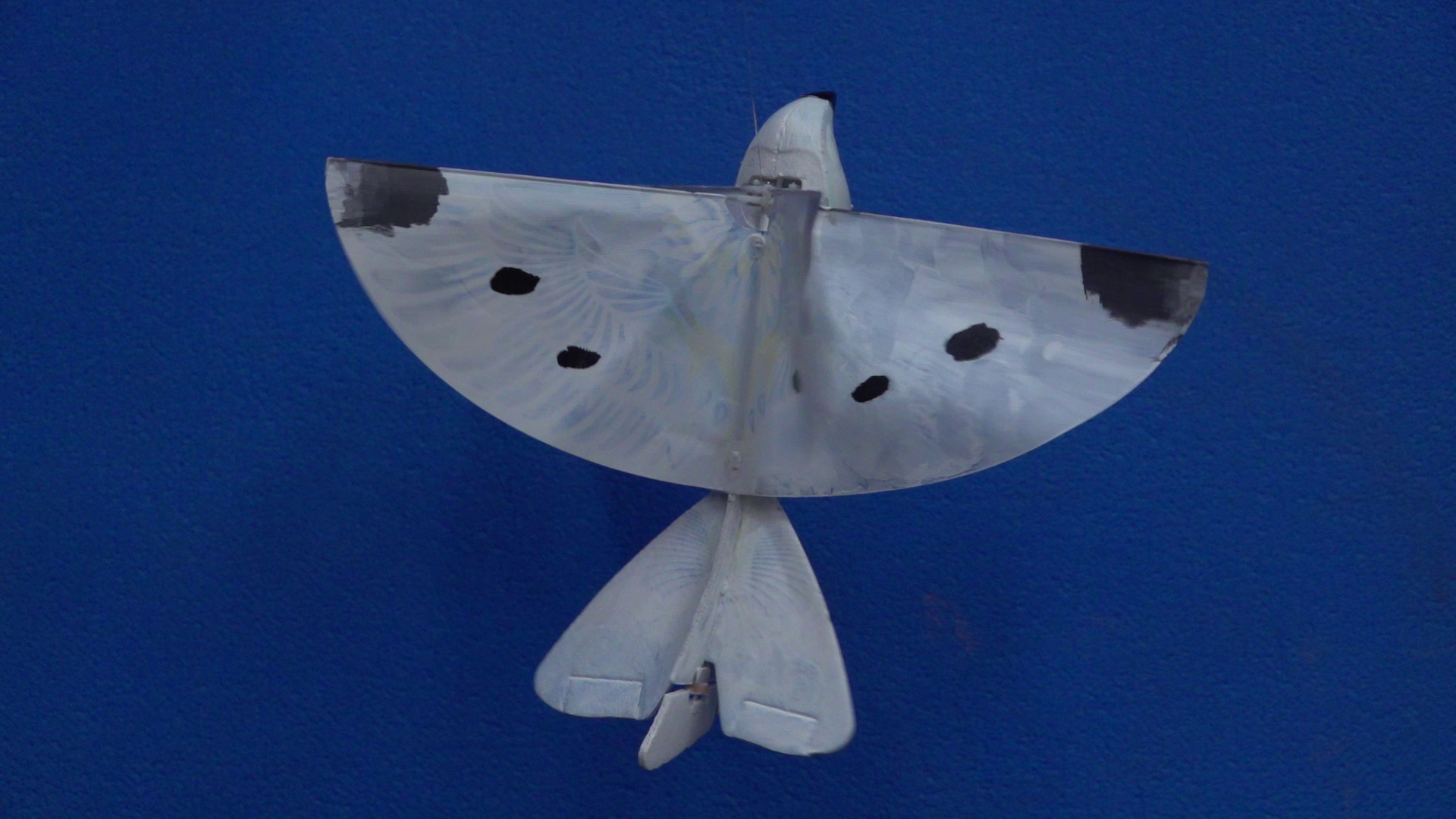}
}
\caption{Real Butterfly (left) and Robot (right)} 
\label{super-white}
\end{figure}
\begin{figure}[h!]
\centering{
\includegraphics[width=1.5in]{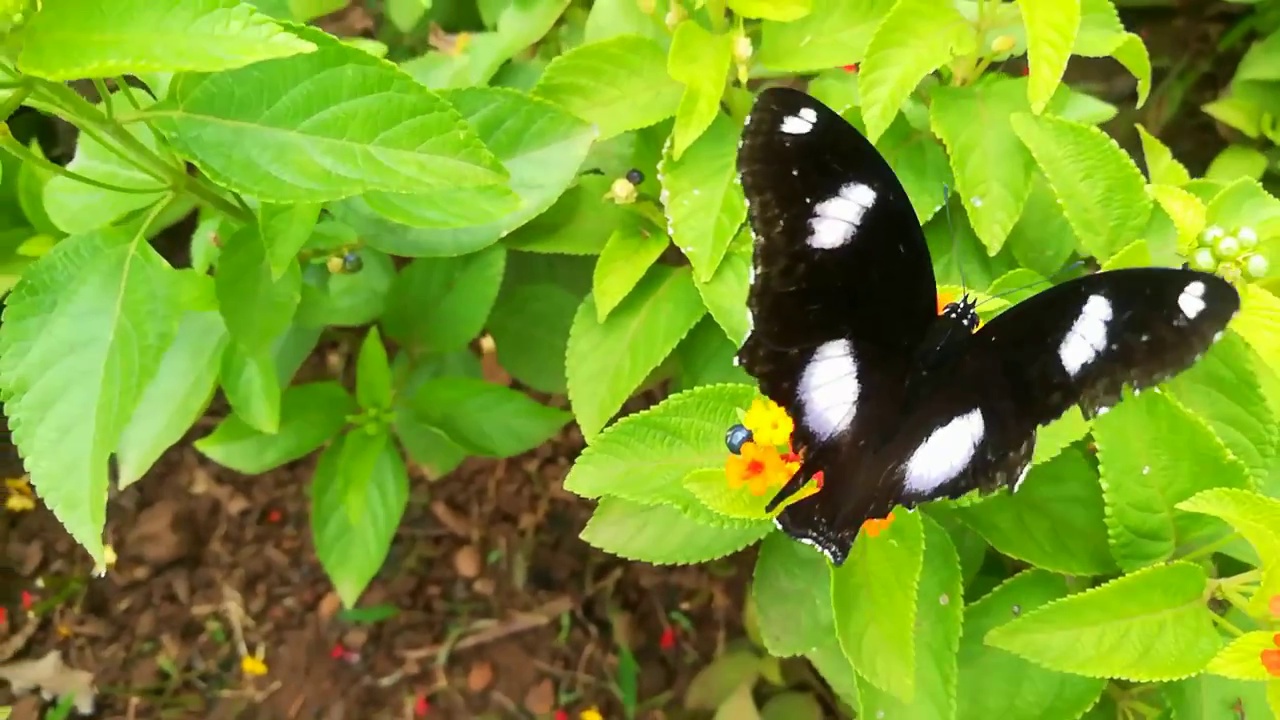} 
\hspace{0.1in}
\includegraphics[width=1.5in]{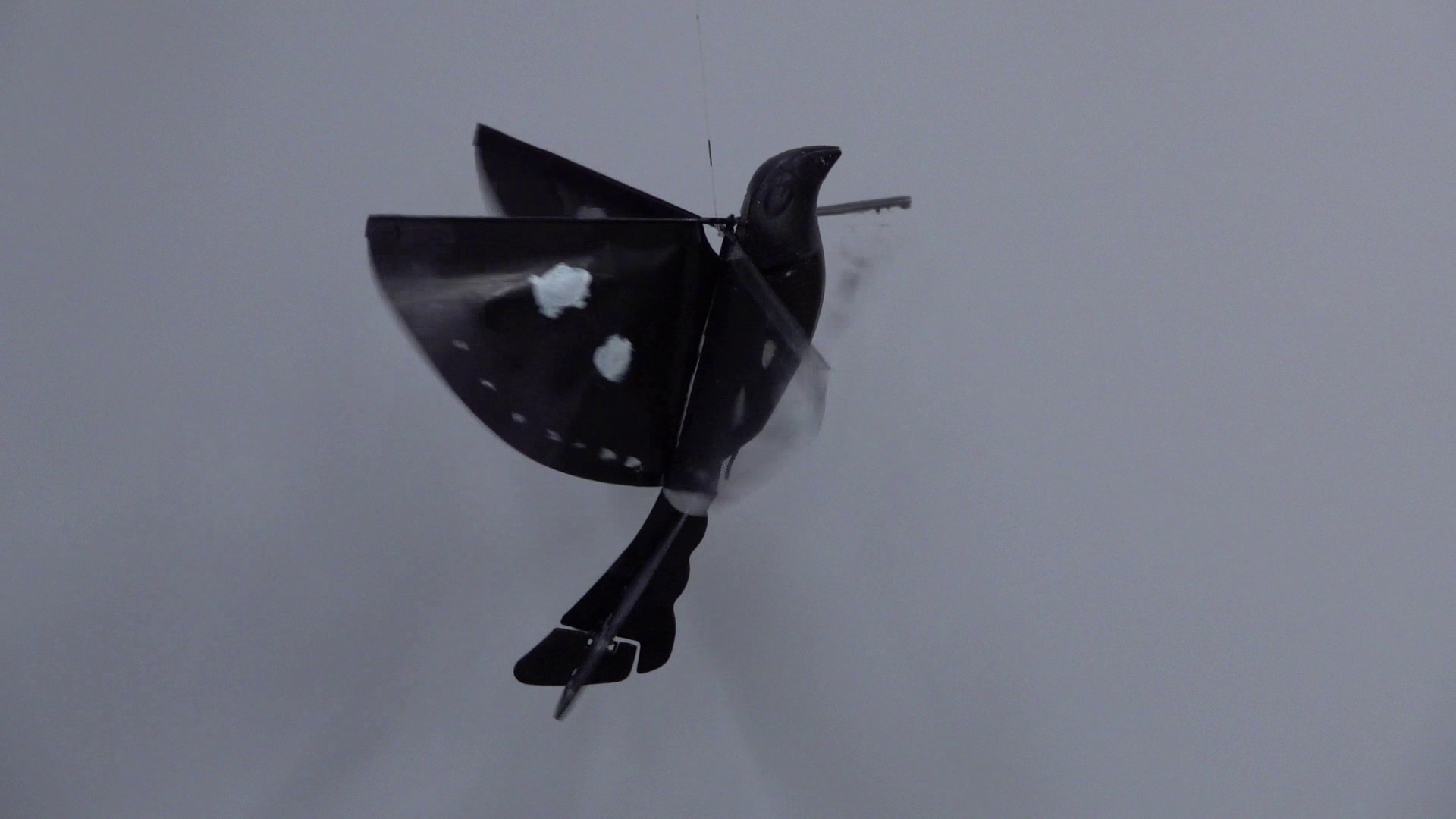}
}
\caption{Real Butterfly (left) and Robot (right)} 
\label{super-black}
\end{figure}

We have three ornithopters (i.e., flapping wing robots) identical in shape. 
The right panels in Figures
~\ref{super-orange}, \ref{super-white}, \ref{super-black} show
the front, the back, and the side of the robots. The original robot is
shaped as a bird, but has two pairs of wings, one
pair with color and the other pair transparent. The
tail is similar to that of a small aircraft. 

We apply an adversarial camouflage to the robots by 
superimposing the patterns copied
from butterflies onto the robots. Figure~\ref{super-white} is the
robot showing the original color design, with only several black
dots added on the wings. Figure~\ref{super-orange} shows the robot with
the body painted black and the wings painted as orange with black
stripes, resembling the pattern on a butterfly.
Figure~\ref{super-black} shows a robot painted as black with a
few white dots on the wings and the tail, resembling a mostly
black butterfly. Hence the three robots in Figures
~\ref{super-orange}, \ref{super-white}, \ref{super-black} all become
a superposition of three different patterns -- the head and the
body resembling a bird, the tail resembling a small aircraft, the
wings resembling a butterfly.

We record them flying by flapping the wings using a Sony DSC-RX10
digital camera with H.264 video encoder to produce mp4 files. 
Both the raw videos and
the frames extracted from the videos are analyzed. The extracted
video frames' resolution is 1920$\times$1080.  
We extract video frames equally spaced in
time and use the state-of-the-art image classification
algorithm to label the selected video frames. First, the video frames are
directly labeled by the pre-trained Inception V3 image
classifier \cite{inceptionv3-2016}, a deep convolutional neural 
network trained on images from ImageNet 1000 classes \cite{imagenet2009}. 
Here we use the
TensorFlow implementation of Inception V3 to label the video
frames \cite{tensorflow}.  

The robots are not an exact match with the image classes
used in the training process for the pre-trained Inception V3. Unsurprisingly, Inception
V3 top one labeled class includes different types of butterflies,
necklace, sweatshirt, crampfish, mask, quill pen, umbrella,
bow-tie etc. None of the video frames are recognized as bird which 
the robots are created to be.  

Next we retrain Inception V3 three times with 9 image
classes \cite{2014transfer}: 1) real bird; 2) real butterfly; 3)
robot; 4) frog; 5) lion; 6) black cat; 7) white rat; 8) fish; 9)
jellyfish. In the robot class, when every time we retrain Inception
V3, we include the frames from two videos, and use the frames
from the third video as the test samples. 

Table~\ref{tab:retrain} shows the classification results from
re-trained Inception V3. In total, the robot with orange pattern has 240
frames; the robot with white wing and black dots has 264 frames;
and the robot with black wing and a few white dots has 203
frames. The majority of the one with orange pattern, 233 frames
out of 240, is labeled as butterfly. The majority of the one with black wing
and white dots, 195 frames out of 203, is labeled as
robot. Surprisingly, the result is split for the one with white
wing -- 182 frames are labeled as butterfly and 82 frames are
labeled as robot.     

\begin{table}{t}
\begin{tabular}{|l|l|}\hline
Orange &  butterfly(233), robot(7) \\ \hline
White  & butterfly(182), robot(82) \\ \hline
Black  & butterfly(1), robot(195), bird(4), black cat(3)\\ \hline
\end{tabular}
\caption{Classification Results from Re-trained Inception V3}
\label{tab:retrain}
\end{table}

\begin{figure}[t!]
\centering{
\includegraphics[width=3.75in]{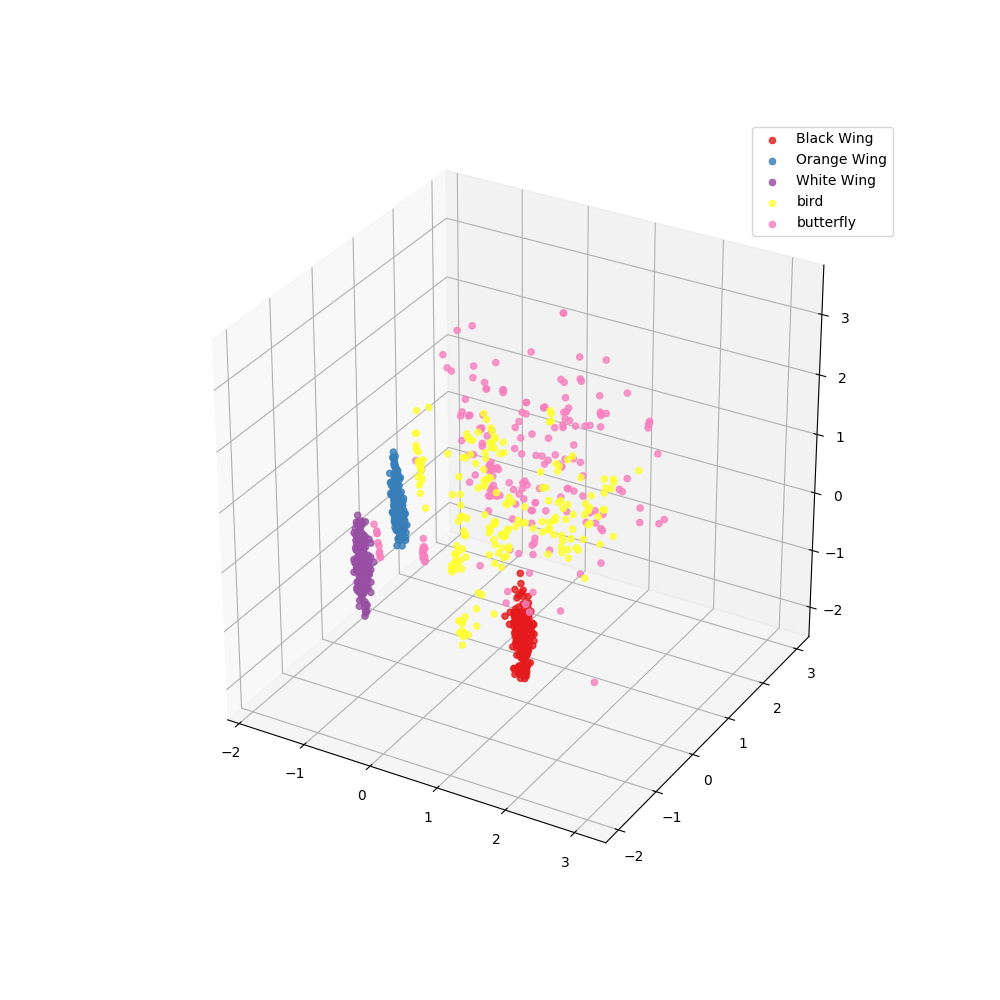}
}
\caption{PCA display of video frames from three robots and images
of real birds and butterflies.   
} 
\label{pca}
\end{figure}

To visualize how the frames from the robots and the images from
other classes are grouped, we perform a principal component
analysis (PCA), including the frames from three robots, and the
images of real bird and real butterfly. Figure~\ref{pca} shows
that the real bird images and the real butterfly images form two
large and loose clusters, whereas the frames from three robots
form three small and tight clusters. 

The results from the re-trained Inception V3 point to a
major vulnerability of DNN --  when several patterns are
superposed on a object, DNN does not know for sure which class the object
belongs to.  We can argue the robot with the orange pattern is
mostly labeled as butterfly because that is a strong identifying
feature, and the black robot is mostly labeled as robot because
the black paint highlights the shape of the robot. For the
white robot, it is the original design with a few added black
dots. Within the tightly clustered frames, 31\% is labeled as
robot and 69\% labeled as butterfly. DNN simply picks one pattern
and assigns the class label. Unfortunately we do not know which
pattern would be picked by DNN. Targeting the current generation of
AI, camouflaging a moving physical object is a successful
adversarial attack, as demonstrated by the robot with orange
pattern. Introducing minor perturbations to a moving physical
object does cause some confusion from DNN, though not as successful
as a full body camouflage, shown by the white and the black robots.  

\noindent
\paragraph{Attack by Motion}

We use the state-of-the-art object detection system, You
only look once (YOLO) V3, to identify the flying robot in
three videos. YOLOv3 \cite{yolov3-2018} is a real time
object detection system. It can identify and label objects from
both images and videos\footnote{https://pjreddie.com/darknet/yolo/}.   
YOLOv3 has a fully convolutional structure with 106 layers. Its 
structure allows it to detect objects of different sizes, from
small to large. If an object is detected in a video, YOLOv3 places
a bounding box around the object with label(s). YOLOv3 is able to  
assign multiple labels to one object if it believes the object fits
the descriptions. YOLOv3 can detect up to 9000 object classes. 
Note object detection, among other applications, is one of the most
important tasks that must be done properly by autonomous
driving systems to avoid collisions. 

The video output from YOLOv3 can be downloaded
here\footnote{https://www.stat.purdue.edu/$\sim$xbw/bio-yolov3/}. 
In two videos, YOLOv3 is ``blind'' -- it cannot detect any object 
at all. In one video
with the mostly black robot, YOLOv3 briefly identifies the tail as two
remotes with probabilities 0.71 and 0.61. Two bounding boxes
focus on the white dots on the tail. This happens during a brief
period that the black robot is 
flying with its back steadily facing the camera.  

We believe the flapping wing motion likely reduces the dependency
among the video frames. Hence the flapping wing motion completely
fools the object detector YOLOv3 in two videos, and succeeds most
of the time in the third video. During our experiment, we notice
there exists other similar motion to reduce the dependency among
video frames and make the object detector ``blind''.  For
example, a man rolls a kayak with a paddle. As the kayak floats
down the river and with the motion of the paddle, YOLOv3 cannot
detect the man. 

Certain motion, when designed properly, is able to fool the
state-of-the-art object detector. Motion should be a
critical part of a physical adversarial attack. Meanwhile we realize that
developing algorithms to pair certain motion with a given object
may be a more difficult task than developing attacks by
adding minor perturbations.    

\begin{figure}[tbh!]
\centering{
\includegraphics[width=3in]{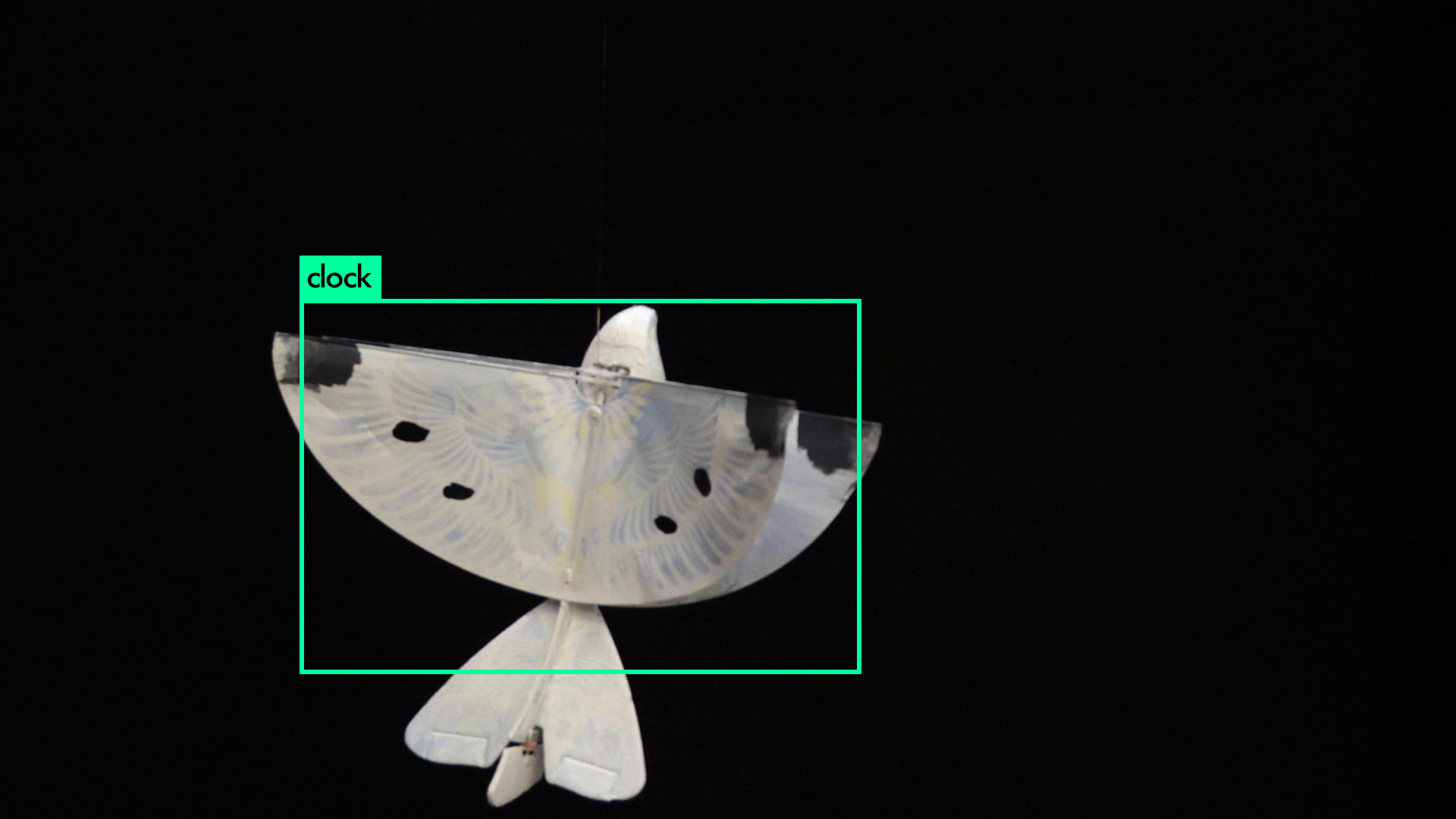}
}
\caption{Class with the highest
score is Clock 0.58, by YOLOv3 trained on ImageNet 9000 classes.
} 
\label{aug-yolov3}
\end{figure}

When using YOLOv3 to label the extracted frames, it places a bounding box
around an object. Whereas Inception V3 provides only
probabilities for the top classes, the bounding box from YOLOv3
allows us to examine why YOLOv3 makes a mistake with a
frame/image. Figure~\ref{aug-yolov3} shows a labeled frame.  
YOLOv3 only detects part of the wings of the white robot
and labels that portion as a clock with score 0.58, due to the black
dots painted on the wings. We notice the bounding
box focuses on the black dots. Depending on which image classes
are used in the training process of a DNN based AI, it
chooses one of the matching classes and assigns it to the frame. Our
experiments demonstrate that DNN cannot tell which is the correct
object class among several matching classes. It is far less accurate than human
recognition facing a complex object.

\section{Conclusion}
\label{sec:conclusion}

Based on the success of the experiment, we demonstrate that it is
possible to design bio-inspired adversarial attack using moving 
physical objects. Future work
includes a targeted attack to create the most effective
camouflage. We believe for an attack to be most powerful, it
should not simply compute a physical outfit
targeting one specific AI. While attackers can launch
unpredictable attacks to break an AI, defender can
update and implement newer and more powerful learning
algorithm at unpredictable times as well. Hence an effective physical
attack must be able to break any AI systems, both present and
future ones, and allow both moving and stationary objects to
avoid being detected. An adversarial camouflage could be 
based on the surrounding environment.  If an object 
operates in a complex environment, its physical
appearance should allow it to melt into the background. If an
object operates in a simple environment, its
physical appearance can be designed to mimic a living organism.     

At the same time, we realize there are objects with radically
different appearances falling in one class, and there are 
hidden objects that can only be detected based on the subtle
irregularities of its shape. AI needs to improve its capabilities,
beyond simply increasing the size of its training data, to
properly recognize and label such objects.   

In our experiment, we use convolutional neural network (CNN)
based YOLOv3 to detect the flying robots 
in the videos. YOLOv3 is not able to detect anything in the
entire video, i.e., bounding box not present throughout the
video. The flapping wing motion makes many frames
blurry, and the object keeps turning and showing different side
of its body in front of the camera. It may be the case that it is
more difficult to locate a dense region in a blurry
image/frame. Also a rapidly moving object makes it difficult to
match the frames with a ground truth object. For future work, we
could see if motion detection or frame control during the
preprocessing stage could help to mitigate the threat of certain
motions. 

Many other state-of-the-art object detection systems are
also based on the convolutional structure, such as
R-FCN~\cite{dai2016rfcn} and
RetinaNet \cite{lin2018focal}. Meanwhile currently most of the adversarial
attacks and defenses target CNN based systems. It would be
interesting to investigate whether Recurrent Neural Network
(RNN) based vision systems are more robust to rapid motions,
since RNN can go deep in the time domain. Note both attacking and
defending RNN need to take into account of the sequential
nature of the data \cite{attack-rnn-2016,defense-rnn-2019}. Hence
it is a different line of research compared with attacking and defending
CNN based systems.

\section{Acknowledgments}
This paper is partially supported by Northrop Grumman Corp. and
SAMSI GDRR 2019-2020 program. 

\bibliographystyle{aaai}
\bibliography{bio.bib}

\begin{thebibliography}{}

\bibitem[\protect\citeauthoryear{Athalye \bgroup et al\mbox.\egroup
  }{2018}]{3d-turtle-2018}
Athalye, A.; Engstrom, L.; Ilyas, A.; and Kwok, K.
\newblock 2018.
\newblock Synthesizing robust adversarial examples.
\newblock In {\em Proceedings of the 35th International Conference on Machine
  Learning},  284--293.

\bibitem[\protect\citeauthoryear{Biggio and Roli}{2018}]{advl-10year-2018}
Biggio, B., and Roli, F.
\newblock 2018.
\newblock Wild patterns: Ten years after the rise of adversarial machine
  learning.
\newblock {\em Pattern Recognition} 84:317--331.

\bibitem[\protect\citeauthoryear{Carlini and
  Wagner}{2017}]{attack-carliniL2-2017}
Carlini, N., and Wagner, D.
\newblock 2017.
\newblock Towards evaluating the robustness of neural networks.
\newblock In {\em 2017 IEEE Symposium on Security and Privacy (SP)},  39--57.

\bibitem[\protect\citeauthoryear{Dai \bgroup et al\mbox.\egroup
  }{2016}]{dai2016rfcn}
Dai, J.; Li, Y.; He, K.; and Sun, J.
\newblock 2016.
\newblock R-fcn: Object detection via region-based fully convolutional
  networks.

\bibitem[\protect\citeauthoryear{Deng \bgroup et al\mbox.\egroup
  }{2009}]{imagenet2009}
Deng, J.; Dong, W.; Socher, R.; Li, L.-J.; Li, K.; and Fei-Fei, L.
\newblock 2009.
\newblock Imagenet: A large-scale hierarchical image database.
\newblock In {\em 2009 IEEE conference on computer vision and pattern
  recognition},  248--255.

\bibitem[\protect\citeauthoryear{Eykholt \bgroup et al\mbox.\egroup
  }{2018}]{stop-sign-2018}
Eykholt, K.; Evtimov, I.; Fernandes, E.; Li, B.; Rahmati, A.; Xiao, C.;
  Prakash, A.; Kohno, T.; and Song, D.
\newblock 2018.
\newblock Robust physical-world attacks on deep learning visual classification.
\newblock In {\em The IEEE Conference on Computer Vision and Pattern
  Recognition (CVPR)},  1625--1634.

\bibitem[\protect\citeauthoryear{Floreano and Mattiussi}{2008}]{bioAI}
Floreano, D., and Mattiussi, C.
\newblock 2008.
\newblock {\em Bio-inspired artificial intelligence: theories, methods, and
  technologies}.
\newblock MIT press.

\bibitem[\protect\citeauthoryear{Goodfellow, Shlens, and
  Szegedy}{2015}]{advl-train-2015}
Goodfellow, I.~J.; Shlens, J.; and Szegedy, C.
\newblock 2015.
\newblock Explaining and harnessing adversarial examples.
\newblock In {\em International Conference on Learning Representations},
  1--12.

\bibitem[\protect\citeauthoryear{Kurakin \bgroup et al\mbox.\egroup
  }{2018}]{nips2017competition}
Kurakin, A.; Goodfellow, I.; Bengio, S.; Dong, Y.; Liao, F.; Liang, M.; Pang,
  T.; Zhu, J.; Hu, X.; Xie, C.; et~al.
\newblock 2018.
\newblock Adversarial attacks and defences competition.
\newblock In {\em The NIPS'17 Competition: Building Intelligent Systems}.
  Springer.
\newblock  195--231.

\bibitem[\protect\citeauthoryear{Kurakin, Goodfellow, and
  Bengio}{2017}]{attack-BIM-physical-2017}
Kurakin, A.; Goodfellow, I.; and Bengio, S.
\newblock 2017.
\newblock Adversarial examples in the physical world.
\newblock In {\em Proceedings of the 6th International Conference on Learning
  Representations (ICLR)},  1--10.

\bibitem[\protect\citeauthoryear{Lin \bgroup et al\mbox.\egroup
  }{2018}]{lin2018focal}
Lin, T.-Y.; Goyal, P.; Girshick, R.; He, K.; and Dollar, P.
\newblock 2018.
\newblock Focal loss for dense object detection.
\newblock {\em IEEE Transactions on Pattern Analysis and Machine Intelligence}.

\bibitem[\protect\citeauthoryear{Papernot \bgroup et al\mbox.\egroup
  }{2016}]{attack-rnn-2016}
Papernot, N.; McDaniel, P.~D.; Swami, A.; and Harang, R.
\newblock 2016.
\newblock Crafting adversarial input sequences for recurrent neural networks.
\newblock In {\em Proceedings of the 35th IEEE Military Communications
  Conference (MILCOM 2016)},  49--54.

\bibitem[\protect\citeauthoryear{Redmon and Farhadi}{2018}]{yolov3-2018}
Redmon, J., and Farhadi, A.
\newblock 2018.
\newblock Yolov3: An incremental improvement.
\newblock {\em arXiv preprint arXiv:1804.02767}.

\bibitem[\protect\citeauthoryear{Rosenberg \bgroup et al\mbox.\egroup
  }{2019}]{defense-rnn-2019}
Rosenberg, I.; Shabtai, A.; Elovici, Y.; and Rokach, L.
\newblock 2019.
\newblock Defense methods against adversarial examples for recurrent neural
  networks.

\bibitem[\protect\citeauthoryear{Shao, Zhu, and Li}{2014}]{2014transfer}
Shao, L.; Zhu, F.; and Li, X.
\newblock 2014.
\newblock Transfer learning for visual categorization: A survey.
\newblock {\em IEEE transactions on neural networks and learning systems}
  26(5):1019--1034.

\bibitem[\protect\citeauthoryear{Sharif \bgroup et al\mbox.\egroup
  }{2016}]{attack-face-2016}
Sharif, M.; Bhagavatula, S.; Bauer, L.; and Reiter, M.~K.
\newblock 2016.
\newblock Accessorize to a crime: Real and stealthy attacks on state-of-the-art
  face recognition.
\newblock In {\em Proceedings of the 2016 ACM SIGSAC Conference on Computer and
  Communications Security},  1528--1540.

\bibitem[\protect\citeauthoryear{Szegedy \bgroup et al\mbox.\egroup
  }{2016}]{inceptionv3-2016}
Szegedy, C.; Vanhoucke, V.; Ioffe, S.; Shlens, J.; and Wojna, Z.
\newblock 2016.
\newblock Rethinking the inception architecture for computer vision.
\newblock In {\em Proceedings of the IEEE conference on computer vision and
  pattern recognition},  2818--2826.

\bibitem[\protect\citeauthoryear{{TensorFlow Github Directory}}{}]{tensorflow}
{TensorFlow Github Directory}.
\newblock
  \url{https://github.com/tensorflow/models/tree/master/research/inception}.

\bibitem[\protect\citeauthoryear{Yuan \bgroup et al\mbox.\egroup
  }{2019}]{survey-attack-defense-DNN-2019}
Yuan, X.; He, P.; Zhu, Q.; and Li, X.
\newblock 2019.
\newblock Adversarial examples: Attacks and defenses for deep learning.
\newblock {\em IEEE transactions on neural networks and learning systems}
  1--20.

\end{thebibliography}

\end{document}